\documentstyle[aps,prb,epsfig]{revtex}
\newcommand{\be}{\begin{equation}}
\newcommand{\ee}{\end{equation}}
\begin{document}

\twocolumn[\hsize\textwidth\columnwidth\hsize\csname
@twocolumnfalse\endcsname

\title{Critical point in the strong field magnetoresistance\\
of a normal conductor/perfect insulator/perfect conductor composite\\
with a random columnar microstructure}

\author{David J. Bergman}
\address{School of
Physics and Astronomy, Raymond and Beverly Sackler Faculty of Exact
Sciences\\Tel Aviv University,
Tel Aviv 69978, Israel\\and\\Department of Physics, The Ohio State University,
Columbus, OH 43210-1106}

\date{\today}

\maketitle

\begin{abstract}
A recently developed self-consistent effective medium approximation,
for composites with a columnar microstructure, is applied to such a 
three-constituent mixture of isotropic
normal conductor, perfect insulator, and perfect conductor,
where a strong magnetic field {\bf B} is present in the plane
perpendicular to the columnar axis.
When the insulating and perfectly conducting constituents do not
percolate in that
plane, the microstructure-induced in-plane magnetoresistance
is found to saturate for large {\bf B}, if the
volume fraction of the perfect conductor $p_S$ is greater than that of the
perfect insulator $p_I$. By contrast, if $p_S<p_I$, that
magnetoresistance keeps increasing as ${\bf B}^2$ without
ever saturating. This abrupt change in the macroscopic response, which
occurs when $p_S=p_I$, is a critical point, with the associated
critical exponents and scaling behavior that are characteristic
of such points. The physical reasons for the singular behavior
of the macroscopic response are discussed. A new type of percolation
process is apparently involved in this phenomenon.

\end{abstract}
\vskip1.5pc]

\vspace{-1.45 cm}

\noindent




Theoretical and experimental studies, performed since 1993,
have shown that the macroscopic electrical response of
three-dimensional (3D) composites with a two-dimensional (2D)
or columnar microstructure can exhibit surprising forms of
behavior when subject to a strong magnetic field {\bf B}.
For example, when a periodic array of parallel cylindrical holes
is etched into an otherwise homogeneous free-electron-conductor host,
the system exhibits a strong dependence of the macroscopic or bulk 
effective resistivity on both the magnitude and the direction of
{\bf B}, when {\bf B} and the average current density
$\langle{\bf J}\rangle$ both lie in the plane perpendicular to
the columnar axis. This occurs whenever the Hall-to-Ohmic
resistivity ratio of the conducting host,
$H\equiv\rho_{\rm Hall}/\rho_{\rm Ohmic}=\mu|{\bf B}|=
\omega_c\tau$,
(here $\mu$ is the Hall mobility, $\omega_c$ is the
cyclotron frequency, and $\tau$ is the conductivity relaxation time)
is greater than 1, and appears even if the array has a high
point symmetry, e.g., square or triangular.
\cite{BergStrel94_1,Tornow96,BergStrelDualityPRB99}
A strongly anisotropic magnetoresistance was also found in calculations
on a periodic columnar array of perfectly conducting inclusions
embedded in a similar host. \cite{BergStrelDualityPRB99}
More recently, composites with a disordered columnar microstructure
were considered, using an appropriate modification of the Bruggeman
self-consistent-effective-medium-approximation.
\cite{BergStrelEMAPRB99}
In such a system, if the pure constituents have an
isotropic transport behavior and the microstructure
is isotropic in the plane perpendicular
to the columnar axis, and if {\bf B} lies in that plane,
then both the longitudinal and the in-plane-transverse
components of the bulk effective resistivity tensor
$\hat\rho_e$, denoted by
$\rho^{(e)}_\parallel$ (i.e., $\langle{\bf J}\rangle\parallel
\langle{\bf E}\rangle\parallel{\bf B}$) and
$\widetilde\rho^{(e)}_\perp$, (i.e., $\langle{\bf J}\rangle\parallel
\langle{\bf E}\rangle$ and they both
lie in that plane but are perpendicular
to {\bf B}) respectively, are independent of the direction of
{\bf B} or $\langle{\bf J}\rangle$ in that plane.
In that study, it was found that for a two-constituent
metal/insulator ($M/I$) mixture of this kind, both of these resistivity
components increase as $H^2$ for $|H|\gg 1$, without ever
reaching saturation. This behavior is not confined to the
case where $M$ represents a free-electron-conductor. All
that is required is that its transport behavior be isotropic,
and that the Hall-to-transverse-Ohmic resistivity ratio $H$
have a magnitude that is much greater than 1. An experimental
consequence of this nonsaturating behavior would be that the
bulk effective magnetoresistivities of such an $M/I$ mixture
would continue to increase as ${\bf B}^2$ when {\bf B} is
large enough so that the Ohmic resistivities of the
conducting constituent are saturated but its Hall resistivity
continues to increase as {\bf B}.
By contrast, in a normal conductor/perfect
conductor ($M/S$) mixture of this kind, both of those resistivity
components saturate at finite values when $|H|\gg 1$.
\cite{BergStrelEMAPRB99}

Here we consider the macroscopic response of a three-constituent
columnar composite of isotropic normal conductor, perfect insulator,
and perfect conductor ($M/I/S$), where the 2D microstructure in
the plane perpendicular to the columnar axis is again random,
and a strong magnetic field {\bf B} is applied in that plane.

We use the ``columnar unambiguous self consistent effective
medium approximation (CUSEMA)'', which was developed in
Ref.\ \onlinecite{BergStrelEMAPRB99}. In this approximation,
the self consistency requirement is that the in-plane components 
of the extra electric field {\bf E} produced by isolated
circular-cylindrical inclusions of the different constituents,
embedded in the fictitious uniform effective medium, vanish on average.
A similar requirement is not imposed upon the columnar component
of {\bf E}, which would be unmeasurable in a thin film realization
of such microstructures. Also, the columnar component of
$\langle{\bf J}\rangle$ is always assumed to vanish. These
requirements lead to the following self consistencey equations
(we take $x$ to be the columnar axis)
\be
0=\left\langle\left[(\hat I-\hat\rho_e/\hat\rho_{\rm inc})\cdot
\hat\gamma_{\rm inc}(\hat\rho_{\rm inc},\hat\rho_e)\cdot\hat\rho_e
\right]_{\{yz\}}\right\rangle,\label{CUSEMA}
\ee
where $\hat I$ is the unit tensor,
$\langle\;\rangle$ indicates an average over the different
types of inclusions, while the subscript $\{yz\}$ indicates that
only the $y,z$ components of the $3\times 3$ tensor appearing
in the square brackets $[\;]$ are included in this calculation,
i.e., the $2\times 2$ matrix in the lower right corner of
$[(\hat I-\hat\rho_e/\hat\rho_{\rm inc})\cdot
\hat\gamma_{\rm inc}(\hat\rho_{\rm inc},\hat\rho_e)\cdot\hat\rho_e]$.
The $3\times 3$ tensor $\hat\gamma_{\rm inc}(\hat\rho_{\rm inc},\hat\rho_e)$
gives the uniform local electric field ${\bf E}_{\rm int}$,
which appears inside an isolated circular-cylindrical inclusion, with
resistivity tensor $\hat\rho_{\rm inc}$, inside the otherwise
uniform effective medium host $\hat\rho_e$, when a uniform
electric field ${\bf E}_0$ is applied at large distances
\be
{\bf E}_{\rm int}=\hat\gamma_{\rm inc}\cdot{\bf E}_0.
\label{gamma_inc}
\ee
This tensor must be calculated from $\hat\rho_{\rm inc}$ and
$\hat\rho_e$---this was done for the relevant types of inclusions
in the Appendix of Ref.\ \onlinecite{BergStrelEMAPRB99}.

It is important to note that the self consistency requirements used
to derive CUSEMA differ from the requirements which are used
to set up the conventional two-dimensional, Bruggeman-type,
self-consistent effective medium approximation (SEMA) in the
$y,z$ plane of this system (see Ref.\ \onlinecite{StroudPRB75}
for a discussion of SEMA in the presence of a magnetic field,
when the conductivity tensors are nonscalar tensors). 
Consequently, it should come as no surprise
if the results of those approximations are sometimes quite
different, as shown in Ref.\ \onlinecite{BergStrelEMAPRB99}.
The great advantage of using CUSEMA is that the exact relations which
$\hat\rho_e$ must satisfy as a result of the duality symmetry are
satisfied automatically, whereas they can be seriously violated in
the two-dimensional SEMA. \cite{BergStrelEMAPRB99}

For the resistivity tensors of the effective medium host and the three
different types of inclusions we write
\begin{eqnarray}
\hat\rho_e&=&\rho_M\left(\begin{array}{ccc}
\alpha_e & -\beta_e & 0 \\
\beta_e & \gamma_e & 0 \\
0 & 0 & \lambda_e\end{array}\right);\;\;\;
\hat\rho_M=\rho_M\left(\begin{array}{ccc}
1 & -H & 0 \\
H & 1 & 0 \\
0 & 0 & \nu\end{array}\right);\nonumber\\
&&\label{rho_e_M}\\
\hat\rho_I&=&\rho_I\hat I,\;\;\rho_I\gg\rho_M;\;\;\;
\hat\rho_S=\rho_S\hat I,\;\;\rho_S\ll\rho_M.\label{rho_I_S}
\end{eqnarray}
These forms mean that a magnetic field {\bf B} is applied
along the $z$ axis, which is perpendicular to the columnar
axis $x$.

The form assumed for $\hat\rho_M$ means that the $M$ constituent
exhibits isotropic transport behavior even when {\bf B} is very
strong, and does not rule out a dependence of its transverse
and longitudinal Ohmic resistivities, $\rho_M$ and $\nu\rho_M$, and
also of its Hall coefficient $H\rho_M/|{\bf B}|$, upon the
{\em magnitude} of {\bf B}. However, it does rule out the possibility
of clean single crystal inclusions of a transition metal like Copper,
where the resistivities become very sensitive to the crystal
orientation with respect to {\bf B} and {\bf J} at strong fields
and low temperatures due to the existence of open orbits on
its Fermi surface, although polycrystalline inclusions of such
a metal might still be allowed. By contrast, the form assumed for
$\hat\rho_e$ means that we expect the bulk effective transport
behavior to be at least mildly
anisotropic as a result of the columnar microstructure,
i.e., in general we expect to have $\alpha_e\neq\gamma_e$.

The forms
assumed for $\hat\rho_I$ and $\hat\rho_S$ include the case where
$\hat\rho_I=\infty$ and $\hat\rho_S=0$. In fact, it is only for
those extreme values that a mathematical singularity will
actually be found to appear in the bulk effective macroscopic response,
justifying the term ``critical point''.

The same resistivity scale $\rho_M$ is used in both $\hat\rho_M$
and $\hat\rho_e$, whose in-plane components will obviously have
similar magnitudes when neither the $I$ constituent nor the $S$
constituent percolates in that plane, i.e.,
when their volume fractions $p_I$, $p_S$ satisfy $p_I<1/2$,
$p_S<1/2$. However, because there
are perfectly conducting inclusions that span the system from
end to end along the columnar axis $x$, therefore the
components of $\hat\rho_e$ which involve that axis will all
vanish. More precisely, we will have
$\alpha_e=O(\rho_S/\rho_M)$, $\beta_e=O(\rho_S/\rho_M)$ as long
as the volume fraction $p_S$ of the $S$ constituent is nonzero.

The $\hat\gamma_{\rm inc}(\hat\rho_{\rm inc},\hat\rho_e)$ matrices
for the three types of inclusions which must be considered are
\cite{BergStrelEMAPRB99}
(we omit the subscript $e$ from $\alpha_e$, $\beta_e$, $\gamma_e$,
and $\lambda_e$ in these expressions, in order to save space)
\begin{eqnarray}
\lefteqn{\hat\gamma_{\rm inc}(\hat\rho_M,\hat\rho_e)=}\nonumber\\
&=&\left(\begin{array}{ccc}
1 & 0 & 0 \\
{-\frac{\beta}{\alpha}+\frac{\gamma H}{1+H^2}\over
\left(\gamma\over\lambda\right)^{1/2}+
\frac{\gamma}{1+H^2}} & 
{1+\left(\gamma\over\lambda\right)^{1/2}\over
\left(\gamma\over\lambda\right)^{1/2}+
\frac{\gamma}{1+H^2}}& 0 \\
0 & 0 &
{1+\left(\gamma\over\lambda\right)^{1/2}\over
1+\frac{\lambda}{\nu}\left(\gamma\over\lambda\right)^{1/2}}
\end{array}\right),\label{gamma_M}
\end{eqnarray}
where we omitted the term $\beta^2/\alpha$,
which is $O(\rho_S/\rho_M)\ll 1$, when it appeared alongside
$\gamma$, which is $O(1)$,
\be
\hat\gamma_{\rm inc}(\hat\rho_I,\hat\rho_e)=\left(\begin{array}{ccc}
1 & 0 & 0 \\
{-\frac{\beta}{\alpha}\over
\left(\gamma\over\lambda\right)^{1/2}} &
{1+\left(\gamma\over\lambda\right)^{1/2}\over
\left(\gamma\over\lambda\right)^{1/2}}& 0 \\
0 & 0 &
1+\left(\gamma\over\lambda\right)^{1/2}
\end{array}\right),\label{gamma_I}
\ee
where we also discarded $O(\rho_M/\rho_I)\ll 1$ terms, and
\be
\hat\gamma_{\rm inc}(\hat\rho_S,\hat\rho_e)=\left(\begin{array}{ccc}
1 & 0 & 0 \\
{-\frac{\beta}{\alpha}\over
\gamma\frac{\rho_M}{\rho_S}}  &
{1+\left(\gamma\over\lambda\right)^{1/2}\over
\gamma\frac{\rho_M}{\rho_S}} & 0 \\
0 & 0 &
{1+\left(\gamma\over\lambda\right)^{1/2}\over
\frac{\rho_M}{\rho_S}\lambda\left(\gamma\over\lambda\right)^{1/2}}
\end{array}\right),\label{gamma_S}
\ee
where we kept some $O(\rho_S/\rho_M)\ll 1$ terms, because those terms
are subsequently multiplied by an $O(\rho_M/\rho_S)\gg 1$
term---see Eq.\ (\ref{CUSEMA}).

\begin{figure}[h]
\centerline{
\epsfig{figure=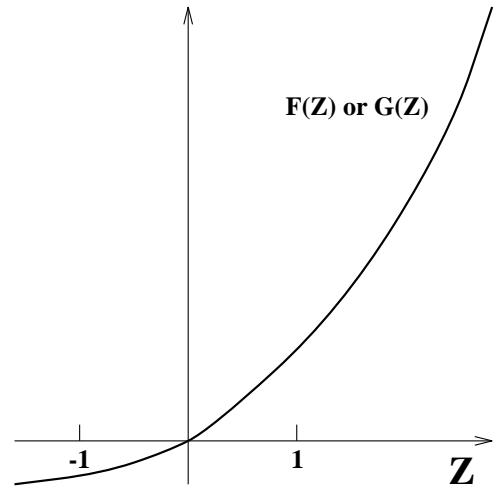,width=2.5 in}
}
\vskip1pc
\caption[]{Qualitative shape of the two scaling functions $F(Z)$
and $G(Z)$---see Eqs.\ (\protect\ref{F_Z}) and (\protect\ref{G_Z}).}
\label{F_Z_G_Z}
\end{figure}

There will be two coupled nontrivial self-consistency equations,
resulting from the $yy$ and $zz$ components of Eq.\
(\ref{CUSEMA}) (the $yz$ and $zy$ components of that equation
are satisfied automatically due to the 2D isotropy of the
microstructure). Thus,
we get two coupled algebraic equations for the two unknown
quantities $\lambda_e\equiv\rho^{(e)}_\parallel/\rho_M$ and
$\gamma_e\equiv\widetilde\rho^{(e)}_\perp/\rho_M$:
\begin{eqnarray}
0&=&\left(\gamma_e\over\lambda_e\right)^{1/2}
\left[(1-2p_S)(1+H^2)-(1-p_I)\gamma_e+p_I\lambda_e\right]\nonumber\\
&&-{\gamma_e\over\lambda_e}p_S(1+H^2)+(2p_I-1)\gamma_e
+(1-p_S)(1+H^2),\nonumber\\
&&\label{yy_eq}
\end{eqnarray}
\be
\left(\gamma_e\over\lambda_e\right)^{1/2}=
\frac{\nu p_S-\gamma_e p_I}{\nu(1-p_S)-\lambda_e(1-p_I)}.\label{zz_eq}
\ee
These equations can be transformed into a single polynomial
equation for, say, $\lambda_e$. That would be an 8th order equation,
which is quite complicated and which I have not been able to
factorize algebraically. On the other hand, if we are interested
only in the asymptotic behavior of $\lambda_e$ and $\gamma_e$ when
$|H|\gg 1$, then an explicit solution can be obtained using
asymptotic analysis. The results are
\begin{eqnarray}
\gamma_e&\cong&\left\{\begin{array}{ll}
H^2\frac{p_I-p_S}{1-2p_I}\frac{1-p_I}{p_I} &
  {\rm for}\;p_I>p_S,\;|(p_I-p_S)H|\gg 1,\\ & \\
\nu\frac{1-2p_S}{p_S-p_I}\frac{1-p_S}{p_S} &
  {\rm for}\;p_I<p_S,\;|(p_I-p_S)H|\gg 1,\\ & \\
\sqrt{\nu}|H|\frac{1-p_I}{p_I} &
  {\rm for}\;p_I\stackrel{>}{\scriptstyle <}p_S,\;|(p_I-p_S)H|\ll 1,
\end{array}\right.\nonumber\\
&&\label{gamma_e}\\
{\gamma_e\over\lambda_e}&\cong&\left\{\begin{array}{ll}
\left(1-p_I\over p_I\right)^2 &
  {\rm for}\;\gamma_e\gg 1,\\ & \\
\left(1-p_S\over p_S\right)^2 &
  {\rm for}\;\gamma_e=O(1).\end{array}\right.\label{gamma_e_lambda_e}
\end{eqnarray}
These results can be recast with the help of two scaling functions
\be
\gamma_e\cong\frac{\nu}{p_I-p_S}F(Z),\;\;\;
\lambda_e\cong\frac{\nu}{p_I-p_S}G(Z),
\ee
where the scaling variable is $Z\equiv|H|(p_I-p_S)/\sqrt{\nu}$, and
the scaling functions have the following limiting forms
\begin{eqnarray}
F(Z)&\cong&\left\{\begin{array}{ll}
\frac{1-p_I}{p_I}\frac{Z^2}{1-2p_I} &
  {\rm for}\;Z\gg 1,\\ & \\
-\frac{1-p_S}{p_S}(1-2p_S) &
  {\rm for}\;Z\ll -1,\\ & \\
\frac{1-p_I}{p_I}Z &
  {\rm for}\;|Z|\ll 1,\end{array}\right.\label{F_Z}\\
G(Z)&\cong&\left\{\begin{array}{ll}
\frac{p_I}{1-p_I}\frac{Z^2}{1-2p_I} &
  {\rm for}\;Z\gg 1,\\ & \\
-\frac{p_S}{1-p_S}(1-2p_S) &
  {\rm for}\;Z\ll -1,\\ & \\
\frac{p_I}{1-p_I}Z &
  {\rm for}\;|Z|\ll 1.\end{array}\right.\label{G_Z}
\end{eqnarray}
Both of these functions have the qualitative form shown in
Fig.\ \ref{F_Z_G_Z}.

Clearly, $|H|=\infty$, $p_S=p_I$ defines a line of critical
points of the macroscopic magnetotransport of such systems:
For $p_S>p_I$, both $\rho^{(e)}_\parallel$ and
$\widetilde\rho^{(e)}_\perp$ saturate when $|H|\rightarrow\infty$,
whereas for $p_S<p_I$ they both keep increasing as $H^2$ for
$|H|\gg 1$. As $p_S\rightarrow p_I$ from below, the coefficients
of the $H^2$ terms tend to zero as $p_I-p_S$, while if
$p_S\rightarrow p_I$ from above the saturated values of
both $\rho^{(e)}_\parallel$ and $\widetilde\rho^{(e)}_\perp$ 
diverge as $1/(p_S-p_I)$. When $p_S=p_I$, then these resistivity
components keep increasing as $|H|$ for $|H|\gg 1$.
In the vicinity of the critical line, where both
$1/|H|$ and $|p_S-p_I|$ are very small, it is easy to
see that $F(Z)\cong G(Z)$ and $\gamma_e\cong\lambda_e$.

Because these results were obtained within the framework of
CUSEMA, some of these behaviors are not expected to be correct
in detail. We do expect that even a more accurate
calculation of asymptotic behavior (i.e., $|H|\gg 1$) will
exhibit saturated behavior for $p_I<p_S$, nonsaturating
$\propto H^2$ behavior for $p_I>p_S$, and nonsaturating $\propto|H|$
behavior for $p_I=p_S$. However, we expect that more accurate
calculations will show that the critical behavior as $p_S\rightarrow p_I$
is not characterized by the simple forms $1/(p_S-p_I)$ or $p_I-p_S$
which were obtained here, but rather by some noninteger values
of the critical exponents. Such calculations are now in progress.

The behavior found in these calculations 
can be understood qualitatively by recalling 
that the in-plane $y,z$ components of $\hat\sigma_M$ are
\be
\frac{1}{\rho_M}\left(\begin{array}{cc}
\frac{1}{1+H^2} & 0\\
0 & \frac{1}{\nu}\end{array}\right).\label{sig_M_2D}
\ee
For $|H|\gg 1$, this represents a very anisotropic 2D conductor
in the $y,z$ plane, with
$\rho_M\sigma_{M\,zz}=1/\nu=O(1)$ and $\rho_M\sigma_{M\,yy}\cong 1/H^2\ll 1$.
In order to get from end to end of the sample, the in-plane
electric current must make its way between different $S$ inclusions by 
flowing through the $M$ host. If there are many more $S$ inclusions
than $I$ inclusions, then straight line $S$-to-$S$ trajectories can
easily be found that are parallel to $z$, allowing the current
to flow through the $M$ constituent {\em only in the $z$ direction}.
Therefore the macroscopic conductivity will depend only on
$\sigma_{M\,zz}$, and the macroscopic resistivity will saturate
when $|H|\gg 1$. In the opposite case, when there are many more
$I$ inclusions than $S$ inclusions, the current will often not be
able to flow even between neighboring $S$ inclusions only along
$z$, but will have to have a nonzero $y$ component in the $M$ constituent.
This is the low conductivity direction, therefore the macroscopic
conductivity will now be determined primarily by
$\sigma_{M\,yy}\cong 1/(\rho_MH^2)$. Threfore the macroscopic resistivity
will not saturate, but keep increasing as $H^2$ forever.

As the relative proportion of the $I$ and $S$ constituents is
varied, there will be a transition from saturating to nonsaturating
$\propto H^2$ behavior, which will have to be abrupt, i.e., it will
be a singular or critical point of the macroscopic response.
This will be reflected by a similarly abrupt change in the detailed
local current distribution, which will have only $z$-parallel
flow lines in the $M$ constituent  for values of $p_I$ below
the transition point value, but will also have nonzero values of $J_y$ in
that constituent when $p_I$ is above that value. Such a transition
constitutes a new type of percolation phenomenon, which we believe
deserves further study, both theoretical and experimental.

Experimental study of the critical point we have discovered could
be done using a doped semiconductor film as the $M$ constituent,
with a random collection of etched perpendicular holes as the
$I$ constituent, and a random collection of perpendicular
columnar inclusions, made of a high conductivity normal metal,
playing the role of the $S$ constituent. We would like to note that
extremely low temperatures or very clean single crystals
would not be required in order to observe this critical point.
What would be necessary is a large contrast at each stage of
the following chain of inequalities
\be
\rho_S\ll\rho_M\ll H^2\rho_M\ll\rho_I.\label{inequalities}
\ee
If Si-doped GaAs is used as the $M$ host, with a negative charge
carrier density of $1.6\times 10^{18}$ cm$^{-3}$ and a mobility
$\mu=2500$ cm$^2/$V\,s at a temperature of 90\,K,
as in the experiment described in Ref.\
\onlinecite{Tornow96}, then a magnetic field of 40\,Tesla
would result in $H=-10$. Such a material would have an Ohmic
resistivity of $1.6\times 10^{-3}\,\Omega$\,cm, about 1000
times greater than that of Copper. Thus, using Copper for the $S$
inclusions and etched holes for the $I$ inclusions, there
should be no difficulty in satisfying all the above
inequalities.

\acknowledgements

This research was supported in part by grants from the
US-Israel Binational Science Foundation, the Israel Science Foundation,
and NSF Grant DMR 97-31511.


\vspace{0.5 cm} 

\begin{references}

\vspace{-0.8 cm}

\bibitem{BergStrel94_1} D. J. Bergman and Y. M. Strelniker,
Phys.\ Rev.\ B {\bf 49}, 16256 (1994).
\bibitem{Tornow96} M. Tornow, D. Weiss, K. v. Klitzing, K. Eberl,
D. J. Bergman, and Y. M. Strelniker, Phys.\ Rev.\ Lett.\ {\bf 77},
147 (1996).
\bibitem{BergStrelDualityPRB99} D. J. Bergman and Y. M. Strelniker,
Phys.\ Rev.\ B {\bf 59}, 2180 (1999).
\bibitem{BergStrelEMAPRB99} D. J. Bergman and Y. M. Strelniker,
Phys.\ Rev.\ B {\bf 60}, 13016 (1999).
\bibitem{StroudPRB75} D. Stroud, Phys.\ Rev.\ B {\bf 12}, 3368 (1975).


\end{references}
\end{document}